\documentclass[10pt,english]{article}
\usepackage{mathptmx}
\usepackage[T1]{fontenc}
\usepackage[latin9]{luainputenc}
\usepackage{babel}
\usepackage{amssymb}
\usepackage{graphicx}
\usepackage{setspace}
\usepackage[pdfusetitle,
 bookmarks=true,bookmarksnumbered=false,bookmarksopen=false,
 breaklinks=false,pdfborder={0 0 1},backref=false,colorlinks=false]
 {hyperref}

\makeatletter
\newcommand{\lyxaddress}[1]{
	\par {\raggedright #1
	\vspace{1.4em}
	\noindent\par}
}

\makeatother

\begin{document}
\title{Causal horizon from quantum fluctuations}
\author{Simone Franchini}
\date{~}
\maketitle

\lyxaddress{\begin{center}
\textit{Sapienza Università di Roma, Piazza Aldo Moro 1, 00185 Roma,
Italy}
\par\end{center}}
\begin{abstract}
We propose a simple model of quantum void where the flow of time is
deduced directly from quantum fluctuations and the consequent particle--antiparticle
creations. Given a certain number of space--like separated pair creation
events, assumed to happen all at the same initial time, we show that
past and future can be foliated into a sequence of adapted manifolds
based on how many events causally influence them. We also give an
explicit construction for the simplest case of one space dimension.

~

\noindent\textit{keywords: relativity, quantum mechanics, time, world
crystal}

~
\end{abstract}
We deduce a kind of time flow directly from quantum fluctuations by
constructing a foliation of the Minkowski diagram (but this argument
could be readily applied to the Penrose diagram \cite{Penrose2006}
as well) into a sequence of disjoint regions, that we interpret as
isochronic and causally ordered. This is done by associating the ``age''
of a spacetime region to the number of initial pair creations that
are causally influencing it. The exact form of these regions is uniquely
determined by the distribution of the initial pair creation events,
that may happen due to stochastic or quantum fluctuations. 

\begin{figure}
\includegraphics[scale=0.29]{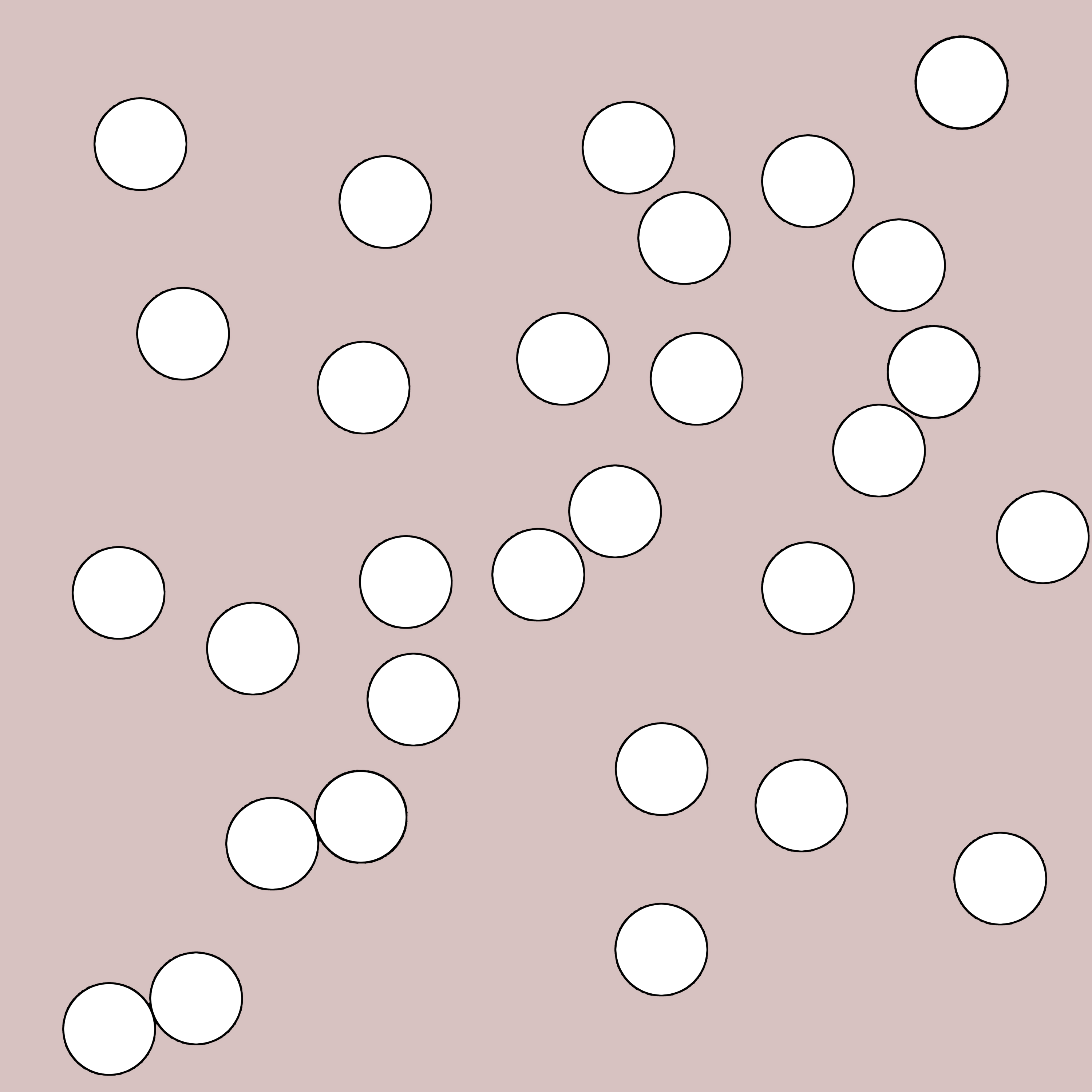}

\caption{\protect\label{fig:Slice-of-spacetime}Isochronic slice of two dimensional
space--time close to the present, $t=0$: the volumes influenced
by one particle appear as a gas of hard spheres, of radius equal to
the elapsed time interval. For small times the areas are mostly disconnected.}
\end{figure}
\begin{figure}
\includegraphics[scale=0.29]{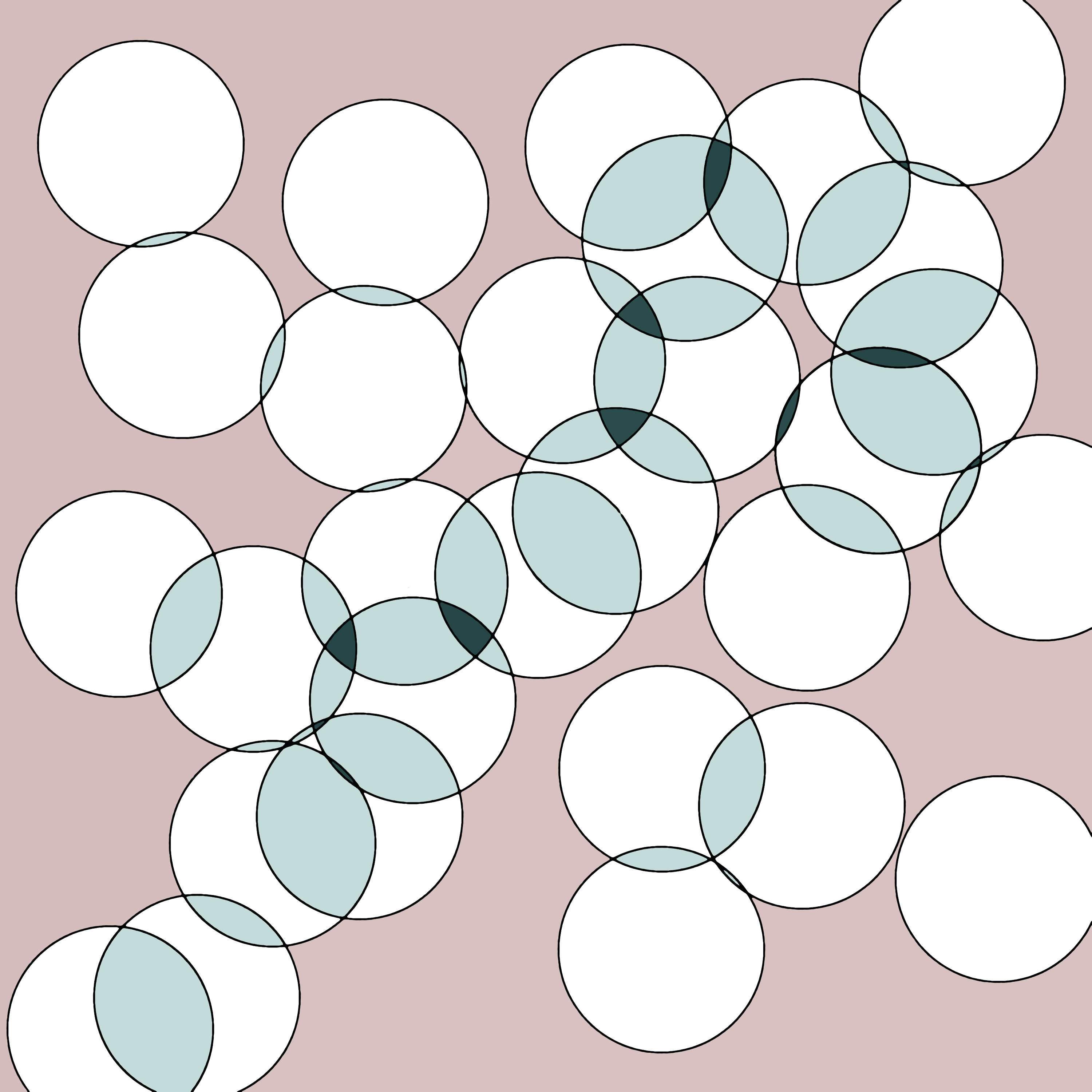}

\caption{\protect\label{fig:After-some-time}After some time the causal fronts
of the observers start to intersect and volumes influenced by more
than one particles appear.}
\end{figure}
\begin{figure}
\includegraphics[scale=0.11]{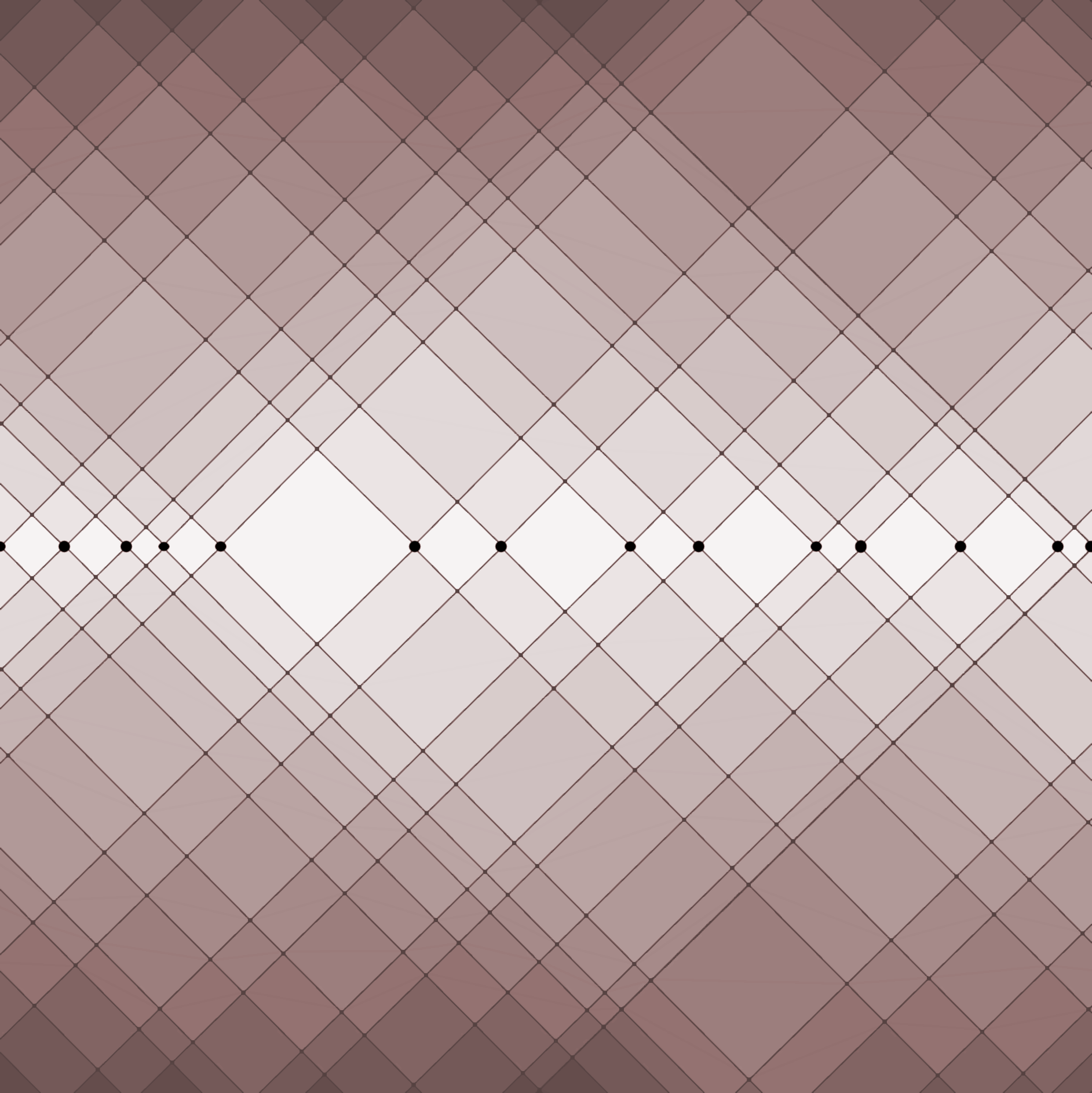}

\caption{\protect\label{fig:Zones-of-causal}Foliation of an $1+1$ Minkowsky
spacetime (completed with its symmetric past) from $N>14$ initial
observers. The space coordinate runs from right to left while time
runs from down to top. For an easier visualization we put $c=1$.
In white are the regions that are causally disconnected from the concentration
points, then in darker shades are the zones influenced by one, two,
three particles and so on.}
\end{figure}
\begin{figure}
\includegraphics[scale=0.11]{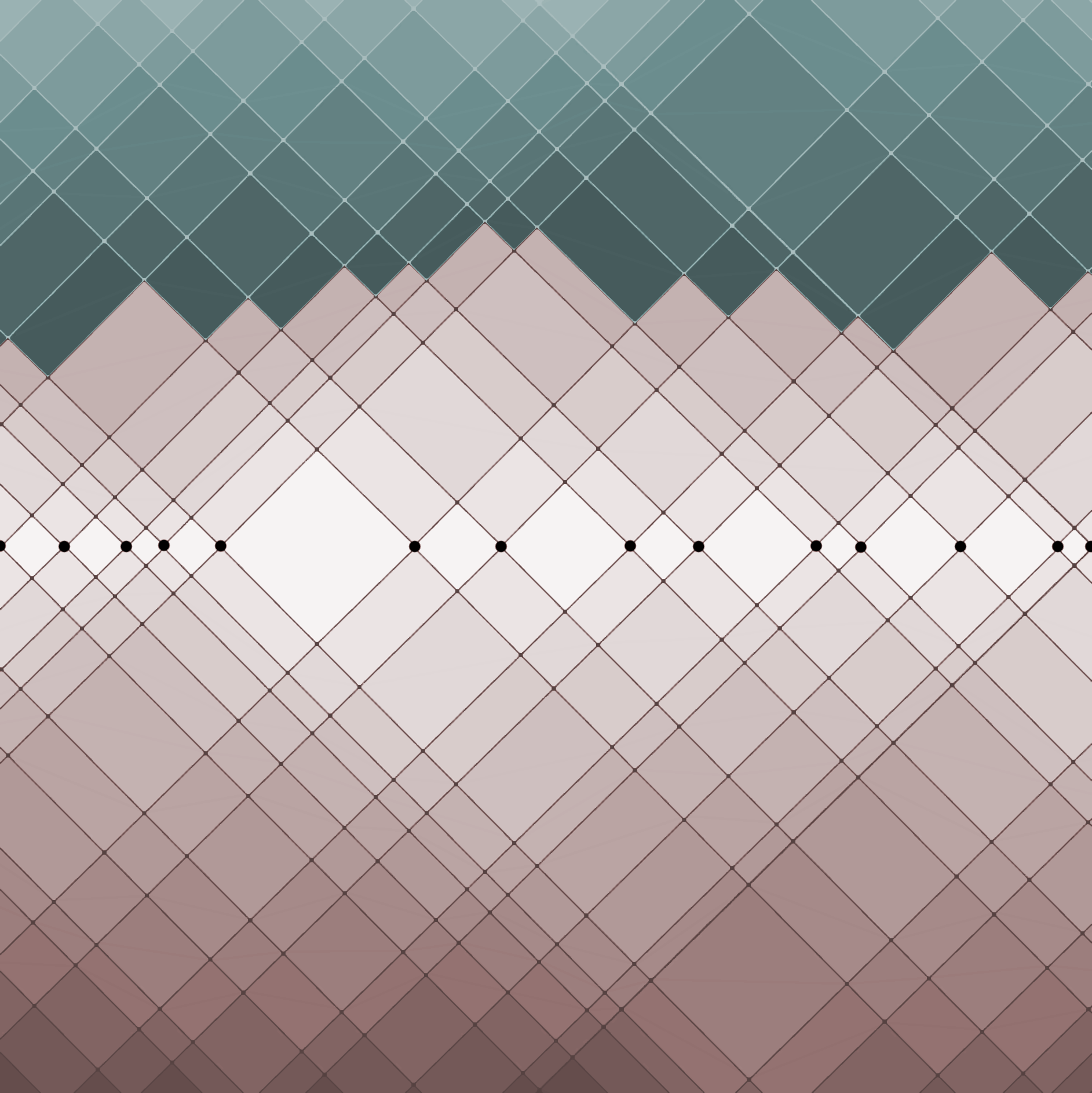}

\caption{\protect\label{fig:Zones-of-causal-2}Foliation of an $1+1$ Minkowsky
spacetime (completed with its symmetric past) from $N>14$ initial
observers. The causal horizon at time $\alpha=5$ is highligted as
the interface between the upper green sector and the lower brighter
sector.}
\end{figure}
\begin{figure}
\includegraphics[scale=0.11]{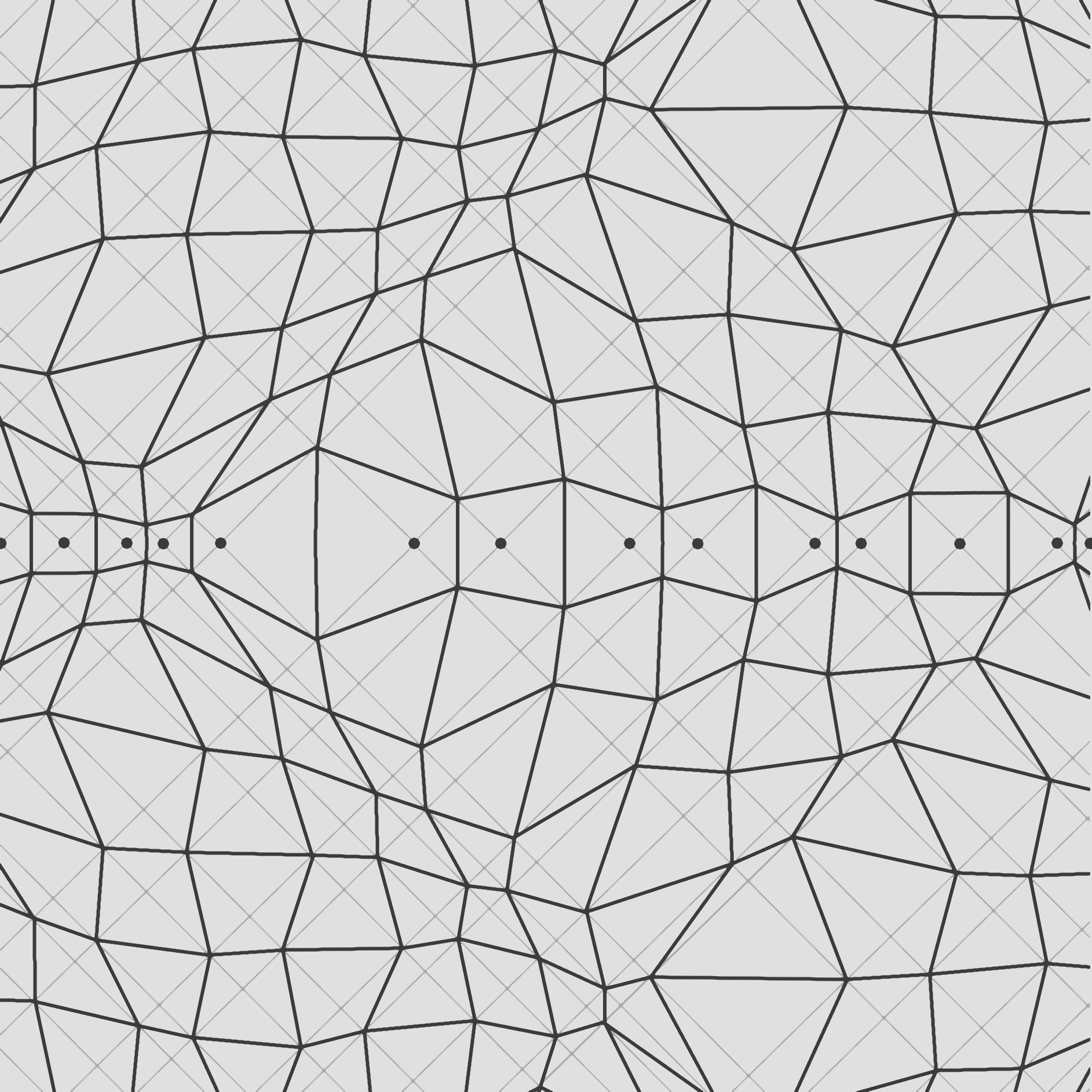}

\caption{\protect\label{fig:Zones-of-causal-1-1}Interpret each crossing as
a center for some deformed Minkowsky diagram}
\end{figure}
\begin{figure}
\includegraphics[scale=0.11]{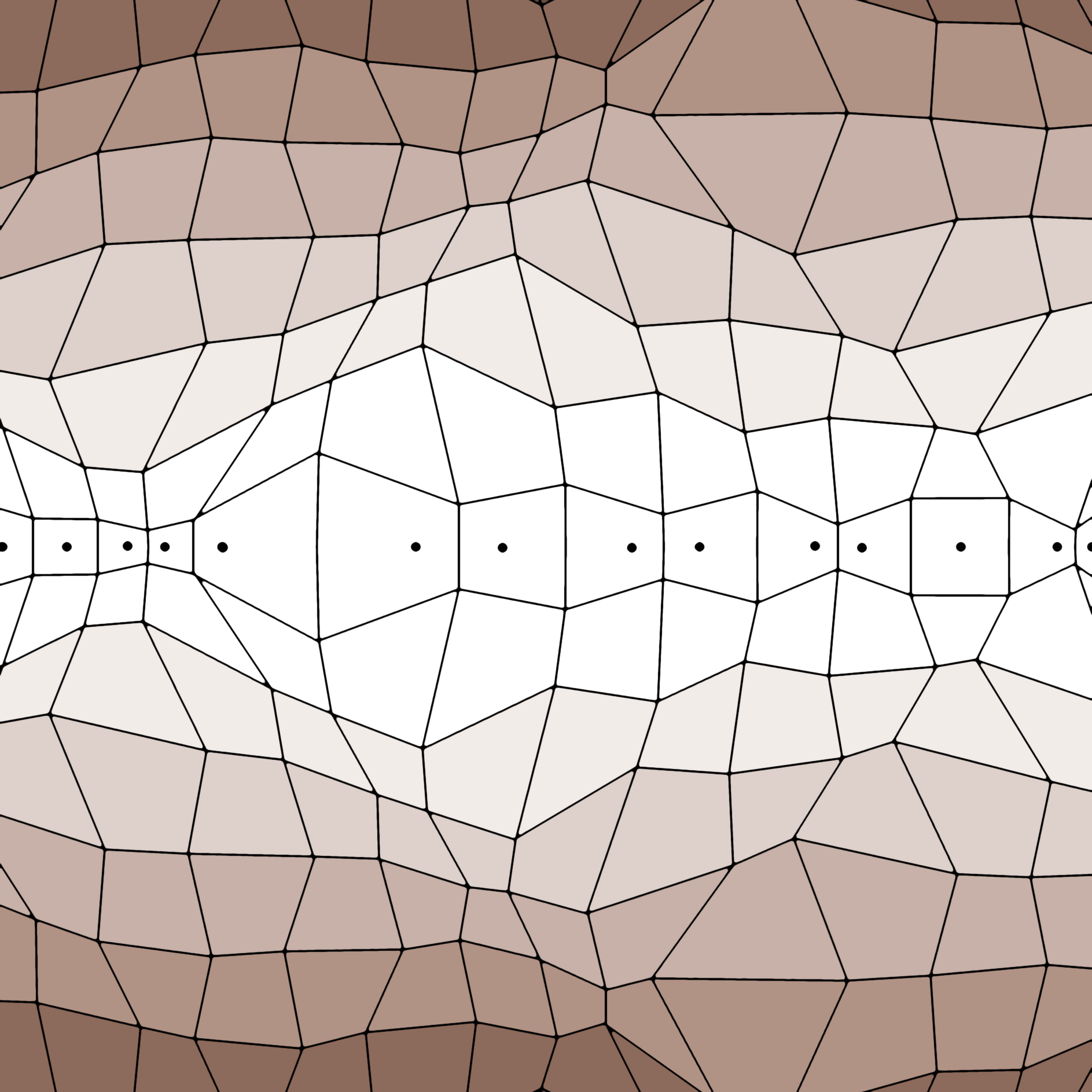}

\caption{\protect\label{fig:Zones-of-causal-1}We can classify the stripes
of spacetime by the number of particles influencing them. The stripes
looks like some kind of interobjective fututure (from center to top,
uncollapsed wavefunction) and intersubjective past (from center to
bottom, collapsed wavefunction) emanating from the central one (stripes
containing the dots, correspond to the present, where the wavefunction
collapse. These intersubjective isocronic stripes could be related
to the classical concept of time.}
\end{figure}

\paragraph{Observers model}

Consider a $d+1$ Minkowsky space--time $\mathcal{M}$ of coordinates
$(x,t)$ where the the $d-$dimensional space component is $x\in\mathbb{R}^{d}$
and $t\in\mathbb{R}$ is the special time. Let now consider a set
of $N$ space points 
\begin{equation}
V:=\{\,x_{i}\in\mathbb{R}^{d}:\,1\leq i\leq N\},
\end{equation}
and suppose that particle--antiparticle pairs are emitted from each
of these points (without creating black holes). We will refer to these
points as observation points, or just observers. A cascade of particles
is generated by each observer as the system start evolving. If we
maintain the speed of light $c$ as an upper limit, these cascades
will be necessarily confined to the light cones originating from the
observation points $x_{i}$. Let us call $\mathcal{M}_{i}^{1}$ the
space--time regions contained in the light cones that emanates from
the points $x_{i}$. For the moment consider only the $i-$th observer
and let 
\begin{equation}
c\,t_{i}\left(x\right)=\left|\,x-x_{i}\right|
\end{equation}
be the light cone emanating from $x_{i}$. The surface $t_{i}\left(x\right)$
is the causal horizon of the observer, as anything outside this cone
will be causally disconnected from $x_{i}$. We can use this definition
to write $\mathcal{M}_{i}^{1}$ in implicit form 
\begin{equation}
\mathcal{M}_{i}^{1}={\textstyle \bigcup_{\ 0\leq t\leq\infty}}\{\left(x,t\right)\in\mathcal{M}:\,t_{i}\left(x\right)\leq t\}.
\end{equation}

In a system with $N$ observers the space--time regions that are
causally connected to the point $x_{i}$ grow in $t$, and after some
time start overlapping with those emanating from the other observers,
depending on their initial distribution (see Figures \ref{fig:Slice-of-spacetime}
and \ref{fig:After-some-time}). This process foliates the Minkowsky
space (or the Penrose diagram as well \cite{Penrose2006}) into zones
of causal influence classified according to how many particles are
are actually influencing them. For example, the union of the space--time
regions influenced by at least one particle is given by the union
of the light cones 
\begin{equation}
\mathcal{M}_{V}^{1}={\textstyle \bigcup_{\ 1\leq i\leq N}}\ \mathcal{M}_{i}^{1},
\end{equation}

The superscript indicates that the surfaces enclose zones that are
influenced by one particle, i.e., whose wave--function is described
by a one body Hamiltonian (no interactions). We can also identify
further zones that are influenced by more than one particle, which
imply higher order of interaction (two--body for zones influenced
by two particles, three--body for those influenced by three and so
on). For example the regions influenced by at least two particles
can be written as follows
\begin{equation}
\mathcal{M}_{V}^{2}={\textstyle \bigcup_{\ 1\leq i_{1}<i_{2}\leq N}}\ \{\,\mathcal{M}_{i_{1}}^{1}\ {\textstyle \bigcap}\ \mathcal{M}_{i_{2}}^{1}\,\},
\end{equation}
and, in general, we can define space--time regions influenced by
at least $\alpha$ particles
\begin{equation}
\mathcal{M}_{V}^{\alpha}={\textstyle \bigcup_{\ 1\leq i_{1}<...<i_{\alpha}\leq N}}\ \{\,\mathcal{M}_{i_{1}}^{1}\ {\textstyle \bigcap}\ ...\ {\textstyle \bigcap}\ \mathcal{M}_{i_{\alpha}}^{1}\,\}.
\end{equation}
The partition is explicitly shown in Figures \ref{fig:Zones-of-causal},
\ref{fig:Zones-of-causal-2}, \ref{fig:Zones-of-causal-1-1} and \ref{fig:Zones-of-causal-1}
for the special case of one space dimension.

\paragraph{Causal horizons}

We identify a sequence of $d-$dimensional surfaces $\partial\mathcal{M}_{V}^{\alpha}$
that separate the regions influenced by $\alpha$ observers from those
influenced by $\alpha-1$ observers. Due to the special nature of
the time coordinate, any of these surfaces can be expressed explicitly
in therm of some time surface $T_{V}^{\alpha}\left(x\right)$ explicit
in $x$. In general we can write $T_{V}^{\alpha}\left(x\right)$ for
the surface $\partial\mathcal{M}_{V}^{\alpha}$ explicitly in $x$,
i.e. 
\begin{equation}
\partial\mathcal{M}_{V}^{\alpha}=\{\left(x,\,T_{V}^{\alpha}\left(x\right)\right)\in\mathcal{M}:\,x\in\mathbb{R}^{d}\}.
\end{equation}

In what follows we will call these surfaces ``causal horizons''.
Of special interest is that with $\alpha=1$ the absolute causal horizon
of the system, i.e. a surface that separate the spacetime regions
influenced by at least one event from those influenced by nothing.
In the case of one particle this time surface is just the light cone
$t_{i}\left(x\right)$, then 
\begin{equation}
\partial\mathcal{M}_{i}^{1}=\{\left(x,\,t_{i}\left(x\right)\right)\in\mathcal{M}:\,x\in\mathbb{R}^{d}\},
\end{equation}
while for many particle the causal horizon $\partial\mathcal{M}_{V}^{1}$
is
\begin{equation}
\partial\mathcal{M}_{V}^{1}=\{\left(x,\,T_{V}^{1}\left(x\right)\right)\in\mathcal{M}:\,x\in\mathbb{R}^{d}\},
\end{equation}
with time surface $T_{V}^{1}\left(x\right)$ computed from $V$ by
solving the variational problem 
\begin{equation}
c\,T_{V}^{1}\left(x\right)=\inf_{1\leq i\leq N}\,c\,t_{i}^{1}\left(x\right)=\inf_{x_{i}\in V}\,\left|\,x-x_{i}\right|.
\end{equation}

\noindent As example, in the Figure \ref{fig:Zones-of-causal} we
show the case with only one space dimension: in this case the picture
is easily visualized because the horizons are just union of segments.
A visualization is still possible also in two and maybe three dimensions,
but the surfaces are much less trivial. In fact, already in two space
dimensions the horizons are composed by pieces of conical surfaces
and can become remarkably complex depending on the initial observers
distribution. However, we notice the connections with the problem
of hard spheres gas \cite{HardSpheres}, for example, if we investigate
what happens near the surface representing the present. Let us consider
a slice of spacetime that is parallel to the present and at distance
$\tau$ in the time dimension (that could well be the Planck time),
like in Figures \ref{fig:Slice-of-spacetime} and \ref{fig:After-some-time}.
The field configuration on this slice, i.e., the hyper--spheres influenced
by a given observer, are mostly disconnected, and the displacement
of their centers will be constrained to be a gas of hard spheres of
radius $\tau$. This analogy offer a direct connection with Spin Glass
and RSB theory trough e.g. \cite{HardSpheres} and references therein.

\paragraph{Isocronic surfaces}

Since for any initial displacement of the observers an absolutely
continuous $d-$dimensional manifold exists separating the region
influenced by at least $\alpha$ observers from that influenced by
less than $\alpha$ observers, then we can identify a partition of
the Minkovsky space into regions influenced by exactly $\alpha$ observers,
i.e., the regions comprised between the causal horizons $\alpha+1$
and $\alpha$: 
\begin{equation}
\mathcal{W}_{V}^{\alpha}:=\mathcal{M}_{V}^{\alpha+1}\setminus\mathcal{M}_{V}^{\alpha}
\end{equation}
Hereafter we identify $t=\alpha$ with the emergent time variable,
the whole region is assumed isochronic in this variable. The region
that encloses the present is identified with the region infuenced
by zero particles:
\begin{equation}
\mathcal{W}_{V}^{0}:=\mathcal{M}_{V}^{1}\setminus\mathcal{M}_{V}^{0}
\end{equation}
Although of limited practical applications, it is interesting to see
how this works in case of just one space dimension, i.e., $1+1$ Minkowsky
spacetime, because the geometry of the partition is simple enough
to be visualized in two dimensions. In this case we can actually draw
the isocronic regions exactly and chart the foliation in a simple
way that can be easily visualized, as in the Figures \ref{fig:Zones-of-causal},
\ref{fig:Zones-of-causal-2}, \ref{fig:Zones-of-causal-1-1} and \ref{fig:Zones-of-causal-1}. 

First of all, notice that, given the restrictions of the one dimensional
space, the horizons are just a sequence of broken lines, that do not
intersect except for several contact points (in common with the previous
and next surface). We identify these as the vertices of our spacetime
lattice \cite{Kleinert1984,Kleinert2004,Lee,Rovelli}. These contact
points, that here we call ``cusps'', can be thought as origins of
a new particle cascade, or a new observer, so that the shape of the
next surface can be obtained from the previous in accordance with
the holographic principle. In one dimension all is straightforward
to visualize: the light cones are one dimensional lines that cut the
Minkowski plane, and when they cross each others the intersection
is just a point. But notice that in more dimensions very intricate
geometries may arise for $d=2$ already. 

For one euclidean dimension the triangulation of the observer at semi-time
$\alpha+1$ associated to the space position $i$ is uniquely determined
by $k=2$ observer at semi--time $\alpha$: the one next on the left
$i-1$ and the other next on the right $i+1$. Then, in this case
it is actually possible to draw the minimal isochronic surface that
separates the $\alpha$--th isochronic region into that influenced
by at least $\alpha$ observers and that influenced by at most $\alpha$
observers: in one dimension these are just broken lines, as it can
be seen in Figures 3 and 4. Notice that adding a new observer will
result in a dylocation in the lattice: this suggest a relation with
the Planck--Kleinert crystal \cite{Kleinert1984,Kleinert2004}. The
resulting random isochronic surfaces can be also interpreted as a
progression of Markov blankets \cite{Judea_Pearl,Friston,Franchini2023,Bardella,BFPF}.
Further investigations are welcome.

\section*{Acknowledgments}

The research presented in this work was conducted in the period 2016--2020
within the LoTGlasSy project (Parisi), funded by the European Research
Council (ERC) under the European Union\textquoteright s Horizon 2020
research and innovation program (Grant Agreement No. 694925). We are
grateful to Giorgio Parisi (Accademia dei Lincei), Sergio Doplicher
(Sapienza Universita di Roma) and Matteo Martinelli (INAF) for interesting
discussions.


\begin{thebibliography}{10}
\bibitem{Penrose2006} \textit{The Road to Reality}, R. Penrose, Vintage
Publishing (2006).

\bibitem{HardSpheres}\textit{Mean-field theory of hard sphere glasses
and jamming}, Parisi, G., Zamponi, F., Rev. Mod. Phys. \textbf{82},
789 (2010).

\bibitem{Kleinert1984} \textit{Gravity as a Theory of Defects in
a Crystal with Only Second Gradient Elasticity}, H. Kleinert, Annalen
der Physik \textbf{499} (2), 117--119 (1987).

\bibitem{Kleinert2004}\textit{Emerging Gravity from Defects in World
Crystal}, H. Kleinert, Brazilian Journal of Physics \textbf{35} (2A),
(2005).

\bibitem{Lee}\textit{Difference equations and conservation laws},
Lee, T. D., J. Stat. Phys. \textbf{46}, 843--860 (1987).

\bibitem{Rovelli}\textit{Discreteness of area and volume in quantum
gravity,} Rovelli, C. Smolin, L., Nucl. Phys. B \textbf{442}, 593--619
(1995).

\bibitem{Judea_Pearl}\textit{Probabilistic Reasoning in Intelligent
Systems: Networks of Plausible Inference}, Pearl, J., In: Representation
and Reasoning Series, San Mateo (1988).

\bibitem{Friston}\textit{The free energy principle made simpler but
not too simple}, Friston, K., Da Costa, L., Sajid, N., Heins, C.,
Ueltzhöffer, K., Pavliotis G.A., Parr., T., Physics Reports \textbf{1024},
1--29 (2023). 

\bibitem{Franchini2023}\textit{Replica Symmetry Breaking without
Replicas}, S. Franchini, Annals of Physics \textbf{450}, 169220 (2023).

\bibitem{Bardella}\textit{Neural Activity in Quarks Language: Lattice
Field Theory for a Network of Real Neurons}, Bardella, G., Franchini,
S., Pan, L., Balzan, R., Ramawat S., Brunamonti E., Pani P., Ferraina
S., Entropy \textbf{26} (6), 495 (2024). 

\bibitem{BFPF}\textit{Lattice physics approaches for Neural Networks},
Bardella, G., Franchini, S., Pani, P., Ferraina S., iScience \textbf{27}
(12), 111390 (2024). 

\end{thebibliography}
\end{document}